\documentclass[twocolumn]{jpsj3}
\usepackage{txfonts}

\title{Superconducting Gap and Symmetry in FeSe$_{1-x}$Te$_x$ Studied by Specific Heat in Magnetic Fields}

\author{Takuya Konno$^1$, Tadashi Adachi$^2$\thanks{E-mail: t-adachi@sophia.ac.jp}, Masato Imaizumi$^1$, Takashi Noji$^1$, Takayuki Kawamata$^1$, and Yoji Koike$^1$}
\inst{$^1$Department of Applied Physics, Graduate School of Engineering, Tohoku University, \\6-6-05 Aoba, Aramaki, Aoba-ku, Sendai 980-8579, Japan \\
$^2$Department of Engineering and Applied Sciences, Faculty of Science and Technology, Sophia University, \\7-1 Kioi-cho, Chiyoda-ku, Tokyo 102-8554, Japan} 

\abst{
In order to investigate details of the superconducting (SC) gap in the iron-chalcogenide superconductors, the specific heat, $C$, of FeSe$_{1-x}$Te$_x$ with $x=0.6-1$ has been measured in magnetic fields. 
Using the two-gap model, it has been found that the smaller SC gap is significantly depressed by the application of magnetic field, resulting in the increase of the slope of the $C/T$ vs $T^2$ plot at low temperatures.  
From the specific-heat measurements at very low temperatures down to 0.4 K, it has been found that the enhancement of the residual electronic-specific-heat-coefficient in the ground state, $\gamma_0$, by the application of magnetic field is much smaller than that expected for superconductors with the typical $s$-wave or $d$-wave SC paring symmetry, which is in sharp contrast to the significant enhancement of $\gamma_0$ observed in the iron-pnictide superconductors. 
These results are discussed in relation to the multi-band effect in the iron-based superconductors.}

\begin{document}
\maketitle

\section{Introduction}

In the research of superconductivity, the understanding of the superconducting (SC) gap provides crucial information on the SC pairing mechanism.
In iron-based superconductors whose SC transition temperature, $T_{\rm c}$, exceeds 50 K, \cite{rA} one of key features is that there exist multiple Fermi surfaces consisting of five orbitals of iron 3d electrons. \cite{r1,r2,r3,r4}
In addition, the existence of at least two kinds of SC gap have been suggested from angle-resolved photoemission spectroscopy (ARPES), \cite{r5,r6} scanning tunneling spectroscopy (STS) \cite{r7} and point-contact Andreev reflection spectroscopy \cite{r8} measurements in both hole-doped (Ba,K)Fe$_2$As$_2$ and electron-doped Ba(Fe,Co)$_2$As$_2$.
As for the SC gap structure, a nodeless gap has been suggested to exist in optimally doped NdFeAsO$_{0.9}$F$_{0.1}$ \cite{rB} and (Ba,K)Fe$_2$As$_2$ \cite{r11} from ARPES, in optimally doped Ba(Fe,Co)$_2$As$_2$ from thermal conductivity \cite{rC} and in LiFeAs from specific-heat \cite{rD} measurements.
On the other hand, a gap with nodes has been suggested to exist in KFe$_2$As$_2$ from ARPES, \cite{r12} in overdoped Ba(Fe,Co)$_2$As$_2$ from specific heat \cite{r15} and in phosphorus-substituted LaFePO \cite{rE} and BaFe$_2$(As,P)$_2$ \cite{rF} from penetration-depth measurements.
Accordingly, in the iron-based superconductors, it is no doubt that there exist at least two kinds of SC gap and that the SC gap structure is different depending on material and carrier doping.

The iron-chalcogenide superconductor FeSe$_{1-x}$Te$_x$ is classified into another group among iron-based superconductors.
Without nominal doping of carriers, $T_{\rm c}$ has been found to increase with increasing $x$ from 8 K at $x=0$, \cite{rG} show the maximum of 14 K at $x=0.6-0.7$ and the superconductivity disappears at $x=1$ where an antiferromagnetic (AF) long-range order appears at low temperatures below $\sim 67$ K. \cite{r17,r18,r19,r20}
Single crystals of FeSe$_{1-x}$Te$_x$ have been grown in a range of $x=0.5-1$, but as-grown single-crystals tend to include excess iron at the interstitial site between the (Se,Te)-(Se,Te) layers in FeSe$_{1-x}$Te$_x$, \cite{r19} resulting in the marked suppression of bulk superconductivity for $x \geq 0.7$. \cite{r22}
Afterward, we have succeeded in obtaining bulk SC single-crystals with a wide range of $x=0.5-0.9$ through the annealing in vacuum ($\sim 10^{-4}$ Pa). \cite{r25}
It has been reported that both the annealing in oxygen for a short time \cite{r29} and the annealing under tellurium vapor \cite{r30} are also effective to obtain bulk SC single-crystals.
These three kinds of annealing are regarded as operating to remove excess iron from the crystals and leading to the appearance of bulk superconductivity. \cite{ohno}

In FeSe$_{1-x}$Te$_x$, two kinds of SC gap have also been reported to exist from specific heat, \cite{r16} muon spin relaxation ($\mu$SR), \cite{rQ,r9} optical conductivity, \cite{r10} penetration depth \cite{cho} measurements.
Moreover, it has been suggested from specific heat, \cite{r22,r27} thermal conductivity, \cite{rH} $\mu$SR \cite{rI} measurements that the SC gap structure of FeSe$_{1-x}$Te$_x$ is nodeless.
Angle-resolved specific heat measurements have suggested that the SC gap is modulated in $k$-space, resulting in the existence of deep minima in the gap. \cite{zeng}
Further detailed measurements of STS \cite{r13} and microwave conductivity \cite{rJ,rK} have claimed that the SC paring symmetry is $s_\pm$-wave corresponding to the spin fluctuation as a glue of electron pairs. \cite{rL,rM}
On the contrary, the study of impurity-substitution effects \cite{rN} has revealed that it is $s_{++}$-wave corresponding to the orbital fluctuation as a glue. \cite{rO,rP}
Therefore, the SC gap in iron-chalcogenide superconductors is controversial even now.

According to specific-heat measurements in magnetic fields by Hu {\it et al}., \cite{r16} the enhancement of the electronic-specific-heat-coefficient, $\gamma$, by the application of magnetic field is smaller than that expected for superconductors with the typical isotropic $s$-wave symmetry, which is interpreted as being most likely due to the multi-band effect.
However, they have estimated the value of $\gamma$ defined as the value of the specific heat, $C$, divided by temperature at a finite temperature of 3 K, so that the value of $\gamma$ is not the residual electronic-specific-heat-coefficient in the ground state, $\gamma_0$.
Therefore, more detailed investigation has been desired.
In this paper, we have performed the specific-heat measurements in magnetic fields for single crystals of FeSe$_{1-x}$Te$_x$ with $x=0.6-1$ at very low temperatures down to 0.4 K in order to clarify details of the SC gap and symmetry.

\section{Experimental Details}

Single crystals of FeSe$_{1-x}$Te$_x$ with $x=0.6-1$ were grown by the Bridgman method. \cite{r25} 
As-grown crystals were annealed at 400 $^{\rm o}$C for 200 h in vacuum ($\sim$ 10$^{-4}$ Pa).
Specific-heat measurements were carried out in magnetic fields of $0 - 9$ T parallel to the c-axis on field cooling by the thermal-relaxation method, using a commercial apparatus (Quantum Design, PPMS).
The Cernox thermometer used for the specific-heat measurements was calibrated in each magnetic-field.
The heat capacity of the crystals was obtained by subtracting the heat capacity of the addenda measured in each magnetic-field.

\section{Results}

\subsection{Specific Heat in Magnetic Fields}
\begin{figure}
\begin{center}
\includegraphics[width=0.9\linewidth]{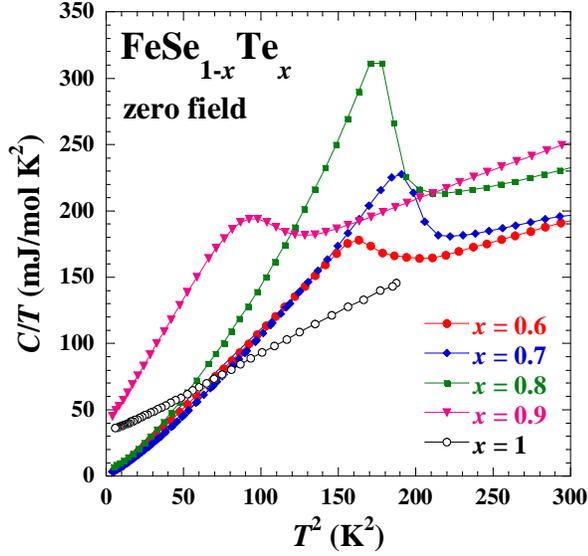}
\caption{(Color online) Temperature dependence of the specific heat, $C$, plotted as $C/T$ vs $T^2$ for FeSe$_{1-x}$Te$_x$ with $x=0.6-1$.  Observed peaks for $x=0.6-0.9$ are due to the superconducting transition.}
\label{f1}
\end{center}
\end{figure}

Figure 1 shows the temperature dependence of the specific heat plotted as $C/T$ vs $T^2$ in zero field for FeSe$_{1-x}$Te$_x$ with $x=0.6-1$.
It is found that a jump of specific heat is clearly observed at $T_{\rm c}$ for $x=0.6-0.9$, indicating the appearance of bulk superconductivity in these crystals.
By contrast, the crystal of $x=1$ shows no SC anomaly, which is consistent with our magnetic-susceptibility results. \cite{r25}
The $T_{\rm c}$ exhibits the maximum at $x=0.7$, while the magnitude of the specific-heat jump at $T_{\rm c}$ exhibits the maximum at $x=0.8$.
The electronic specific-heat, $C_{\rm el}$, in the normal state is found to increase with increasing $x$ at $x=0.6-0.9$, because the phonon specific-heat, $C_{\rm ph}$, does not change with $x$ so much. \cite{r27}
On the other hand, $C_{\rm el}$ is reduced at $x=1$.
These behaviors of $C_{\rm el}$ are explained in terms of the effective mass being enhanced around the boundary between SC and AF phases. \cite{r27}
The value of $\gamma_0$ is nearly zero for $x=0.6-0.8$, suggesting the homogeneous SC state in these crystals.
On the other hand,  $\gamma_0$ is finite for $x=0.9$, suggesting the existence of normal-state carriers even in the SC state. \cite{r27}

\begin{figure}
\begin{center}
\includegraphics[width=0.9\linewidth]{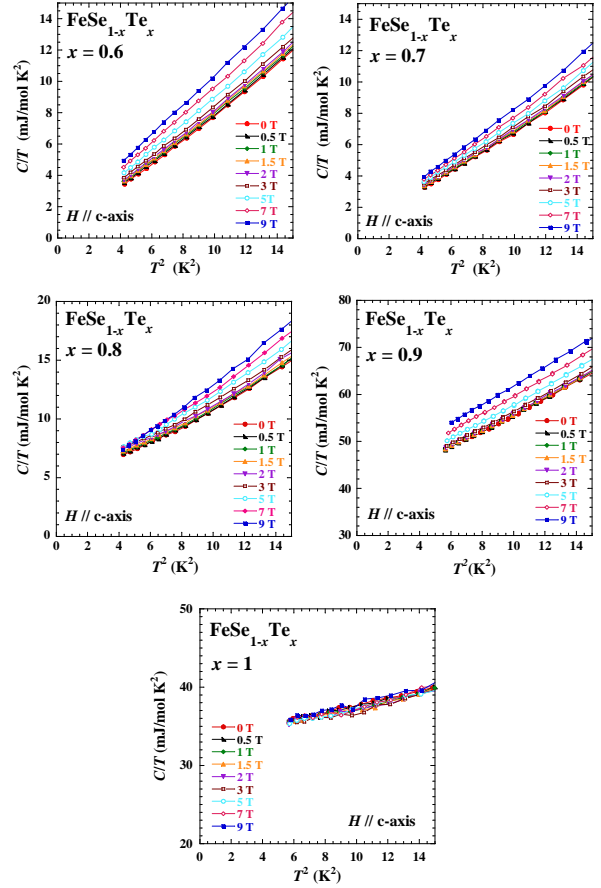}
\caption{(Color online) Temperature dependence of the specific heat, $C$, in various magnetic-fields parallel to the c-axis on field cooling plotted as $C/T$ vs $T^2$ for FeSe$_{1-x}$Te$_x$ with $x=0.6-1$.}
\label{f2}
\end{center}
\end{figure}

The specific heat in various magnetic-fields for $x=0.6-1$ is shown in Fig. 2. 
For $x=1$, the specific heat is almost independent of magnetic field, indicating that $C_{\rm el}$ in the normal state as well as $C_{\rm ph}$ is unaffected by the application of magnetic field.
On the other hand, it is found that not only the value of $C/T$ but also its slope against $T^2$ increases with increasing magnetic-field for $x=0.6-0.8$, as observed in the result of $x=0.57$ by Hu {\it et al}. \cite{r16}
Since the value of $C/T$ is independent of magnetic field for $x=1$, the peculiar behavior of $C/T$ in magnetic fields for $x=0.6-0.8$ is probably due to $C_{\rm el}$ in the SC state.

\subsection{Magnetic-Field-Dependence of Superconducting Gap}

\begin{figure}
\begin{center}
\includegraphics[width=1.0\linewidth]{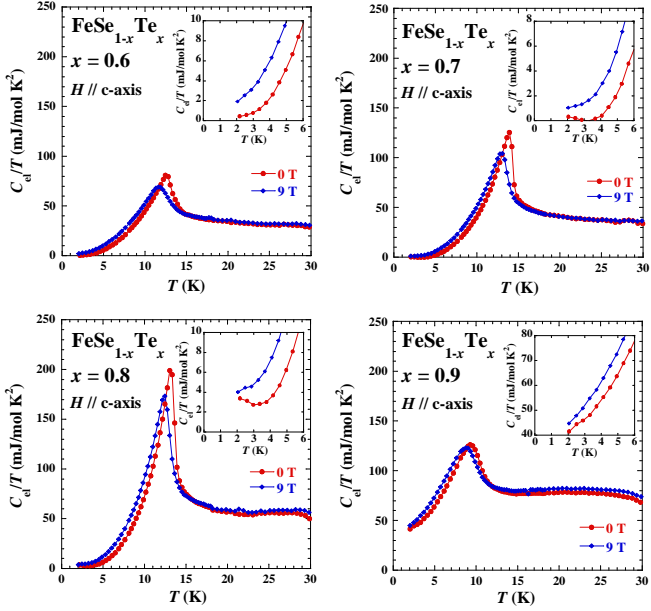}
\caption{(Color online) Temperature dependence of the electronic specific heat divided by temperature, $C_{\rm el}/T$, in zero field and 9 T for FeSe$_{1-x}$Te$_x$ with $x=0.6-0.9$. Insets show magnified plots at low temperatures.}
\label{f3}
\end{center}
\end{figure}

In order to estimate $C_{\rm el}$, $C_{\rm ph}$ was subtracted from the total specific-heat, using the data of $C$ of Fe$_{0.95}$Cu$_{0.05}$Se$_{0.4}$Te$_{0.6}$ in which the superconductivity was completely suppressed through the 5\% substitution of copper for iron.
On the subtraction, the balance of entropy between SC and normal states at the SC onset temperature was taken into consideration precisely.
The details in the estimate of $C_{\rm el}$ have been described in our previous paper. \cite{r27}
Figure 3 shows the temperature dependence of $C_{\rm el}$ divided by $T$ in zero field and 9 T for $x=0.6-0.9$.
It is found that the onset of the SC transition, where $C_{\rm el}/T$ starts to increase quickly with decreasing temperature at $\sim 15$ K for $x=0.6-0.8$ and at $\sim 12$ K for $x=0.9$, is almost unchanged by the application of magnetic field.
In contrast, the position of the SC peak shifts to the low-temperature side and the peak width becomes broad by the magnetic field.
Focusing on the data at low temperatures shown in the insets, the slope of the $C_{\rm el}/T$ vs $T$ plot in 9 T is found to be larger than that in zero field, indicating that the increasing slope of the $C/T$ vs $T^2$ plot by the application of magnetic field shown in Fig. 2 is caused by $C_{\rm el}$, namely, the change of the electronic state in magnetic field.

In order to discuss quantitatively, the SC gap was estimated using the two-gap model.
In this model where two bands are assumed to produce respective two SC gaps independently, the entropy, $S_i$, and the specific heat, $C_i$, of the $i$ ($i=1,2$) band are described as follows, \cite{r34}
\begin{equation}
\frac{S_i}{\gamma_{\rm N \it i} T_{\rm c}} = -\frac{6}{\pi^2} \frac{\it \Delta_i}{k_{\rm B} T_{\rm c}} \int_{0}^{\infty}\Bigl[f_i \ln f_i + \left( 1-f_i \right) \ln \left( 1-f_i \right) \Bigr] dy_i, 
\end{equation}
\begin{equation}
\frac{C_i}{\gamma_{\rm N \it i} T_{\rm c}} = t \frac{d\left( S_i/\gamma_{\rm N \it i} T_{\rm c} \right)}{dt}.
\end{equation}
Here, $\gamma_{\rm N \it i}$ and ${\it \Delta_i}$ are the electronic-specific-heat-coefficient in the normal state and the SC gap of the $i$ band, respectively, and $f_i = \bigl[ \exp \left(\beta E_i \right) +1 \bigr]^{-1}$ and $\beta = \left( k_{\rm B} T \right) ^{-1}$.
The energy of quasiparticles is given by $E_i = \bigl[ \epsilon_i^2 + {\it \Delta}_i^2 \left( t \right) \bigr]^{1/2}$ , where $\epsilon_i$ is the energy of  normal electrons of the $i$ band relative to the Fermi surface and ${\it \Delta_i} \left( t \right) = {\it \Delta_i} \delta \left( t \right)$.
Here, $\delta (t)$ is the normalized BCS gap at the reduced temperature, $t = T/T_{\rm c}$. \cite{r33}
The integration variable in Eq. (1) is $y_i = \epsilon_i / {\it \Delta_i}$.
The total specific-heat is given by the sum of the contributions of each band calculated independently according to Eq. (2), namely, $C = C_1 + C_2$.
The electronic-specific-heat-coefficient in the normal state, $\gamma_{\rm N}$, is given by $\gamma_{\rm N} = \gamma_{\rm N1} + \gamma_{\rm N2}$.
The experimental data of $C$ were fitted with three parameters; ${\it \Delta_{\rm 1}}$, $\it \Delta_{\rm 2}$ and the relative weight, $w \equiv  \gamma_{\rm N1}/\gamma_{\rm N}$.
Namely, $\gamma_{\rm N2}/\gamma_{\rm N} = 1-w$.
In the fitting, the value of $w$ was fixed to be 0.7.
This value was confirmed to give us the almost best-fit results for $x=0.6$ and 0.7.
\begin{figure}
\begin{center}
\includegraphics[width=1.0\linewidth]{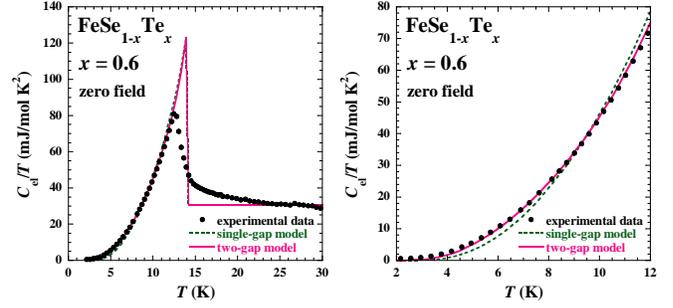}
\caption{(Color online) Temperature dependence of the electronic specific-heat divided by the temperature, $C_{\rm el}/T$, in zero field for FeSe$_{1-x}$Te$_x$ with $x=0.6$ and the best-fit results using single-gap (dashed line) and two-gap (solid line) models.}
\label{f4}
\end{center}
\end{figure}
Figure 4 displays the best-fit result for $x=0.6$ using the two-gap model.
The best-fit result using the single-gap model is also shown, where $i=1$ only in Eq. (2).
Obviously, the experimental data are well reproduced by the two-gap model at low temperatures below the peak temperature.
Therefore, our specific-heat results also prove the existence of at least two SC gaps in FeSe$_{1-x}$Te$_x$.

The dependence on the Te-concentration $x$ of the two SC gaps in zero field and 9 T estimated thus is shown in Fig. 5.
The $\it \Delta_{\rm L}$ and $\it \Delta_{\rm S}$ denote larger and smaller gaps, respectively.
In zero field, it is found that ${\it \Delta_{\rm L}}$ increases with increasing $x$ and shows the maximum at $x=0.8$ where the specific-heat jump shows the maximum as shown in Fig. 1, followed by the rapid decrease toward $x=1$.
On the other hand, the maximum of ${\it \Delta_{\rm S}}$ appears at $x=0.7$ where $T_{\rm c}$ exhibits the maximum.
In a magnetic field of 9 T, intriguing is at $x=0.6-0.8$ that ${\it \Delta_{\rm S}}$ is significantly depressed by the application of magnetic field, whereas the depression of ${\it \Delta_{\rm L}}$ is small.
Accordingly, it is concluded that the significant depression of ${\it \Delta_{\rm S}}$ by the magnetic field leads to excitation of quasiparticles at low temperatures effectively, resulting in the increase of the slope of the $C/T$ - $T^2$ plot with increasing magnetic-field.

\subsection{Magnetic-Field-Induced Change of $\gamma_{0}$ in the Ground State}

\begin{figure}
\begin{center}
\includegraphics[width=0.9\linewidth]{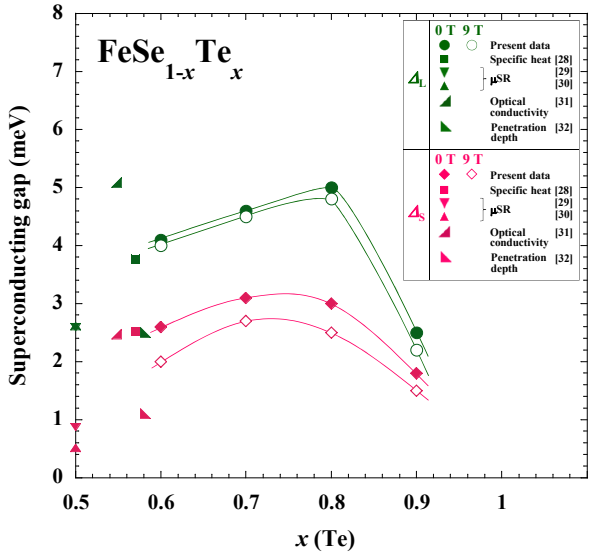}
\caption{(Color online) Te-concentration dependence of larger and smaller superconducting gaps, $\it \Delta_{\rm L}$ and $\it \Delta_{\rm S}$, respectively, obtained using the two-gap model for FeSe$_{1-x}$Te$_x$. Preceding data of ${\it \Delta}_{\it \rm L}$ and ${\it \Delta}_{\it \rm S}$, obtained from specific heat \cite{r16}, $\mu$SR, \cite{rQ,r9} optical conductivity \cite{r10} and penetration depth \cite{cho} measurements, are also plotted for comparison. Solid lines are to guide the reader's eye.}
\label{f5}
\end{center}
\end{figure}

In order to address the issue of the SC paring symmetry, specific-heat measurements were performed in magnetic fields at very low temperature below 2 K where the increase of the slope of the $C/T$ vs $T^2$ plot by the application of magnetic field is expected to be small far below $T_{\rm c}$.
Figure 6 shows the $C/T$ vs $T^2$ plot in various magnetic-fields at very low temperatures down to 0.4 K for $x=0.6$ and 0.7.
The change of the slope of the $C/T$ vs $T^2$ plot is found to become small to some extent at very low temperatures.
It is found that $\gamma_0$ increases with increasing magnetic-field.
It is, however, noted that the increase in $\gamma_0$ is relatively small, compared with the increase in $\gamma$ at 3 K formerly reported for FeSe$_{1-x}$Te$_x$ with $x=0.57$ \cite{r16} and $x=0.5$. \cite{r31}

Figure 7 shows the magnetic-field-dependence of the field-induced residual electronic-specific-heat-coefficient in the ground state, $\Delta\gamma_0 (H) \equiv \gamma_0 (H) - \gamma_0 (0)$, for $x=0.6$ and 0.7.
The horizontal and vertical axes are normalized by the upper critical field, $H_{\rm c2}$, \cite{r35} and $\gamma_{\rm N} - \gamma_0 (0)$, respectively.
As shown in the figure, for a fully gapped $s$-wave superconductor, the field-induced quasiparticle density of states proportional to $\Delta \gamma_0 (H)$ exhibits linear field-dependence, because the quasiparticle density of states is proportional to the number of vortex cores.
On the other hand, in the case of a $d$-wave SC gap with nodes, $\Delta \gamma_0 (H)$ exhibits square-root field-dependence due to the Volovik effect.
However, the present data of $\Delta \gamma_0 (H)$ follow neither $s$-wave nor $d$-wave behavior for $x=0.6$ and 0.7.
Instead, the increase in $\Delta \gamma_0 (H)$ is much smaller than that expected for superconductors with the typical $s$-wave or $d$-wave SC paring symmetry.

\section{Discussion}
From the specific-heat measurements of FeSe$_{1-x}$Te$_x$ with $x=0.6-1$ in magnetic fields, it has been found that the slope of the $C/T$ vs $T^2$ plot at low temperatures increases with increasing magnetic-field for $x=0.6-0.8$ where the specific-heat jump is clearly observed at $\sim 14$ K.
Through the analysis using the two-gap model, it has been concluded that the depression of $\it \Delta_{\rm S}$ by the application of magnetic field leads to the increase of the slope of the $C/T$ vs $T^2$ plot with increasing magnetic-field.
Therefore, in order to discuss the increase of the quasiparticle density of states by the magnetic field, not the value of $\gamma$ estimated at a finite temperature of 3 K \cite{r16,r31} but the value of $\gamma_0$ in the ground state is indispensable.

Using the two-gap model, the magnitudes of two SC gaps, $\it \Delta_{\rm L}$ and $\it \Delta_{\rm S}$, have been unveiled.
In zero field, $\it \Delta_{\rm L}$ shows the maximum at $x=0.8$ where the specific-heat jump shows the maximum, while $\it \Delta_{\rm S}$ shows the maximum at $x=0.7$ where $T_{\rm c}$ shows the maximum.
Specific-heat measurements by Hu {\it et al}. \cite{r16} have revealed that the magnitudes of the two gaps are 3.77 meV and 2.53 meV at $x=0.57$.
The other preceding results have revealed that the two gaps are 2.6 meV and 0.87 meV at $x=0.5$ from $\mu$SR, \cite{r9} 2.61 meV and 0.51 meV at $x=0.5$ from $\mu$SR, \cite{rQ} 5.08 meV and 2.47 meV at $x=0.55$ from optical-conductivity, \cite{r10} 2.5 meV and 1.1 meV at $x=0.58$ from penetration-depth measurements. \cite{cho}
As shown in Fig. 5, the present results are not contradictory to the specific heat and $\mu$SR results, although the optical-conductivity and penetration-depth results show a little large value of $\it \Delta_{\rm L}$ and small values of $\it \Delta_{\rm L}$, $\it \Delta_{\rm S}$, respectively.
The maximum $T_{\rm c}$ not at $x=0.8$ but at 0.7 suggests that the SC transition in FeSe$_{1-x}$Te$_x$ is dominated by $\it \Delta_{\rm S}$.
As shown in Fig. 5, $\it \Delta_{\rm S}$ is suppressed by the application of magnetic field more significantly than $\it \Delta_{\rm L}$ in a wide range of $x=0.6-0.8$ where the homogeneous superconductivity appears.
Since the SC coherence length is anti-correlated with the size of the SC gap, it is reasonable that $\it \Delta_{\rm S}$ tends to be depressed effectively by the magnetic field.
On the other hand, equivalent effects of the magnetic field on the depression of $\it \Delta_{\rm L}$ and $\it \Delta_{\rm S}$ at $x=0.9$ is due to the values of $\it \Delta_{\rm L}$ and $\it \Delta_{\rm S}$ close together, which may be related to the inhomogeneous SC state in real space or $k$-space. \cite{r27}

\begin{figure}
\begin{center}
\includegraphics[width=0.9\linewidth]{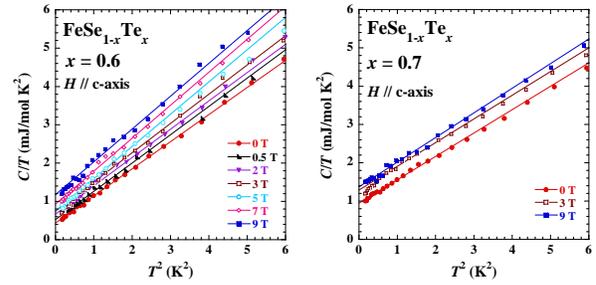}
\caption{(Color online) Temperature dependence of the specific heat, $C$, in various magnetic-fields parallel to the c-axis on field cooling at very low temperatures down to 0.4 K plotted as $C/T$ vs $T^2$ for FeSe$_{1-x}$Te$_x$ with $x=0.6$ and 0.7.}
\label{f1}
\end{center}
\end{figure}

Finally, we discuss the magnetic-field-induced enhancement of $\gamma_0$.
As shown in Fig. 7, the increase in $\gamma_0$ with increasing magnetic-field for $x=0.6$ and 0.7 is much smaller than that expected for superconductors with the typical $s$-wave or $d$-wave SC paring symmetry.
Besides, the increasing behavior of $\gamma_0$ as clearly shown in the inset seems not to be so simple at $x=0.6$.
This unusual field-dependence of $\gamma_0$ is distinct from the behaviors in iron-pnictide superconductors \cite{rD} and from the behavior of the thermal conductivity in FeSe$_{1-x}$Te$_x$ with $x=0$ where the typical multi-gap $s$-wave behavior is observed. \cite{rH}
These results may be related to the theory by Bang \cite{r36} that, for dirty $s_\pm$-wave superconductors with the ratio of two SC gaps being $0.7 - 0.9$ and $\gamma_0 / \gamma_{\rm N}$ in zero field being $\sim 0.22$, the enhancement of $\gamma_0$ by the application of magnetic field is smaller than that expected for superconductors with the simple $s$-wave SC paring symmetry due to the multi-band effect.
Although the ratio of our estimated two gaps is 0.63 and 0.67 for $x=0.6$ and 0.7, respectively, the present enhancement of $\gamma_0$ is too small to be explained by his theory.
As our estimated $\gamma_0 / \gamma_{\rm N}$ for $x=0.6$ is $\sim 0.014$ which is one order of magnitude smaller than that assumed in his theory, our crystals of $x=0.6$ and 0.7 might be in the clean limit at low temperatures. \cite{rK}
According to his theory, moreover, the smaller $\gamma_0 / \gamma_{\rm N}$ becomes, the more $\gamma_0$ is enhanced by the application of magnetic field, which is contradictory to the present result.
Accordingly, further measurements including the specific heat in high magnetic fields to investigate how $\gamma_0$ reaches $\gamma_{\rm N}$ or another theory is necessary to understand the peculiar behavior of $\gamma_0$ in magnetic fields.

\section{Summary}
\begin{figure}
\begin{center}
\includegraphics[width=1.0\linewidth]{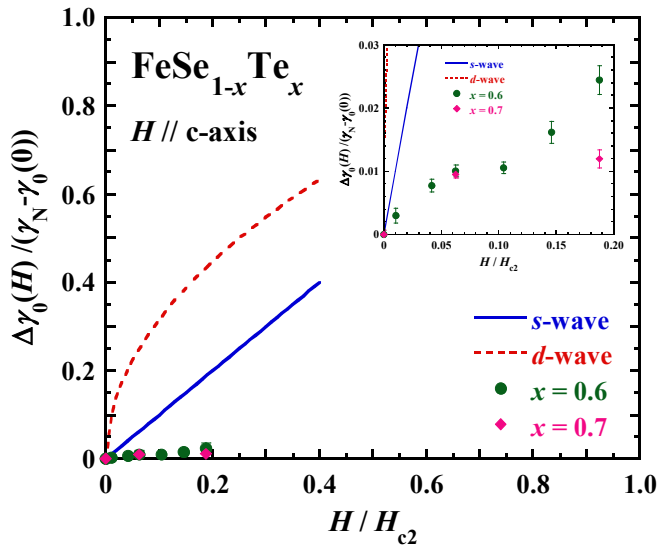}
\caption{(Color online) Magnetic-field-dependence of the field-induced residual electronic-specific-heat-coefficient in the ground state, $\Delta \gamma_0 (H) \equiv \gamma_0 \left( H \right) - \gamma_0 (0)$, normalized by $\gamma_{\rm N} - \gamma_0 (0)$ for FeSe$_{1-x}$Te$_x$ with $x=0.6$ and 0.7. $H / H_{c2}$ represents the reduced field, where $H_{c2}$ = 48 T. \cite{r35} The inset is a magnified plot of Fig. 7 at low fields.}
\label{f7}
\end{center}
\end{figure}

We have investigated magnetic-field effects on the specific heat for FeSe$_{1-x}$Te$_x$ with $x=0.6-1$.
From the temperature dependence of the specific heat plotted as $C/T$ vs $T^2$, it has been found that not only the value of $C/T$ but also its slope increases with increasing magnetic-field.
Using the two-gap model, it has been found that $\it \Delta_{\rm S}$ is significantly depressed by the application of magnetic field, resulting in the increase of the slope of the $C/T$ vs $T^2$ plot at low temperatures with increasing magnetic-field.
On the other hand, $\it \Delta_{\rm L}$ is almost unaffected by the magnetic field.
The specific-heat measurements in magnetic fields at very low temperatures down to 0.4 K have been performed to investigate the change of $\gamma_0$ by the magnetic field.
Surprisingly, the magnetic-field-induced enhancement of $\gamma_0$ is much smaller than that expected for superconductors with the typical $s$-wave or $d$-wave SC paring symmetry.
Although the multi-band effect might be related to the insensitiveness of $\gamma_0$ to magnetic field, \cite{r36} further specific-heat measurements in high magnetic fields are necessary to be conclusive.

\section*{Acknowledgment}

This work was partially supported by JSPS KAKENHI Grant Number 23540399.

\end{document}